\documentclass[pra,twocolumn,a4paper,superscriptaddress,showpacs,prl]{revtex4}
\usepackage{amsmath}
\usepackage{amsfonts}
\usepackage{graphicx}
\begin{document}
\title{Distinctive features in the quantum sub-dynamics of bipartite systems} 
\author{A. K. Rajagopal}
\affiliation{Center for Quantum Studies, George Mason 
University, Fairfax, VA 22030, USA.}
\affiliation{Inspire Institute Inc., McLean, VA 22101, USA.} 
\author{A. R. Usha Devi}
\affiliation{Inspire Institute Inc., McLean, VA 22101, USA.}
\affiliation{Department of Physics, Bangalore University, Bangalore-560 056, India.}
\author{R. W. Rendell}
\affiliation{Center for Quantum Studies, George Mason 
University, Fairfax, VA 22030, USA.}
\author{Michael Steiner}
\affiliation{Center for Quantum Studies, George Mason 
University, Fairfax, VA 22030, USA.}
\affiliation{Inspire Institute Inc., McLean, VA 22101, USA.} 

\date{\today}
\begin{abstract}
Distinctive features in the  sub-dynamics of an interacting bipartite system are explored. The sub-dynamic 
Heisenberg operators are introduced in a novel way leading to the quasi-particle-like concept in the respective %%@
subsystems. This 
formulation is illustrated using the known exact solutions of the generic (Jaynes - Cummings) model of interacting %%@
bipartite system of 
a spin 1/2 particle and a single mode electromagnetic field (more generally, boson field). The striking effects of 
interaction and entanglement are reflected in the dressing of the non-interacting number and spin representations, %%@
displaying 
quasi-particle-like properties. This development may be at the heart of the physics of back - action discussed in the %%@
recent 
literature.  \end{abstract}
\maketitle
In condensed matter physics, the low-lying excitations near the ground state - single particle and collective 
excitations - are described in simple terms using the concept of quasi-particles (QP). They describe the renormalized 
energy and wave-function due to interaction, but retain the pristine nature of the non-interacting particles (e.g., 
fermion nature of the non-interacting electron persists in the QP regime). This concept has been widely used in more 
complex situations than their first recognition in describing interacting electrons and phonons in metals. Many 
important physical properties of the system at low temperatures could be understood in this way. Recently, the QP 
concept has been invoked in exotic systems such as Bose-Einstein condensation in two dimensions, quantum Hall effect, 
and strongly correlated electron and spin systems.  It may also be mentioned that the idea of back-flow 
also became evident in this picture. For a comprehensive review of all these, one may consult~\cite{1}.

	With the advent of recent advances in experimental techniques an entirely different class of problems involving 
time-dependent phenomena, e.g., interacting bipartite system of atoms and radiation~\cite{2,3}, have opened up a new %%@
class of issues. 
Even though the total system evolves in time unitarily, the time evolution of the subsystems 
are quite different~\cite{4}. By introducing sub-dynamic Heisenberg operators, 
we find a novel dynamical description of the subsystem time evolution having 
certain distinctive features, which appear to be similar to QP. 
In view of this observation, for want of a better nomenclature, we suggest 
quasi-particle-like (QPL) for describing these time evolutions. 
The purpose of the present paper is to study the QPL sub-dynamics of both parts of an interacting bipartite system 
on equal footing. In particular, the new Heisenberg operators define QPL, but do not retain the 
pristine nature of the non-interacting particles, except at 
certain intervals of time, unlike the QP. As a byproduct, this formulation enables us to examine 
the complimentary back-action effects, like the back-flow concept mentioned above in the QP context. But 
 unlike in the condensed matter systems, QPL describes time-dependent 
features that occur in certain regions of time. 
These dynamical properties of systems, including back-action, are being pursued by 
various experimental groups around the world.

Quantum-limited solid-state devices have shown that the model of an atom in a cavity, bathed in a laser field, serves %%@
as 
a generic theoretical model for describing these systems.  The Jaynes-Cummings model (JCM)~\cite{5} 
 serves as an exactly solvable model of such two interacting 
disparate systems - a two-level system and an
electromagnetic field (EMF) (or more generally, a boson field) - under certain simplifying assumptions. 
These are the rotating wave approximation and weak coupling of the two-level atom with the field. 
JCM has recently been extended to strong coupling regime~\cite{6}. Inherently, this model has the atom 
entangled with the photon and hence the dynamics of the subsystems of both the photon and the atom 
will exhibit consequent features of dynamical correlation.  Although the sub-dynamics of the photon and spin
(2-state atom) are generally expected to be
complex for the interacting system, the
Heisenberg sub-dynamic operators are shown here to
make explicit the physical processes, which
deviate from the pristine dynamics. This allows
the identification of QPL regimes resembling the
pristine dynamics, with renormalized values for
photon number and spin. 
These features emerge as novel consequences of the formalism presented here. 
It is worth pointing out the use of JCM in several other situations in the recent literature: 
entangled systems for controlling symmetric qubits in trapped ions~\cite{7}, 
linear optics methods to generate symmetric qubits~\cite{8}, collective atomic spin excitations~\cite{9}, 
and stability of atomic clocks~\cite{10}.

We first give a brief description of the theory of sub-dynamics in terms of the Kraus operators for a typical pair of 
systems~\cite{4}. We then construct the Heisenberg representations of typical operators in the two subsystems. We 
apply this formalism to JCM and discuss how this relates to QPL features.  
This study leads to further implications in more general situations.

Consider two systems A and B with the Hamiltonian (time-independent in this sequel)
\begin{equation}
\label{1}	 
H=H_A+H_B+H_{AB}	
\end{equation}
as a generic interacting composite system. The time evolution of  the density matrix of this system, $\rho(t)$, is %%@
expressed in terms of the unitary evolution 
operator, $U(t)=e^{-i\, t\, H}$ , in the standard way~\cite{4}, 
$\rho(t)=U(t)\, \rho(t=0)\, U^\dag(t)$. If $O_S$  is a typical operator in the composite space in the Schrödinger 
representation,  its Heisenberg representation is defined by:  
${\rm Tr}[O_S\rho(t)]={\rm Tr}[O_H(t)\rho(t=0)],\ O_H(t)=U^\dag(t)\, O_S\, U(t).$ 
Here `Tr' stands for the trace over the whole system. 
When the initial density matrix is of the un-correlated form,
$\rho(t=0)=\rho_A(0)\otimes\rho_B(0),$ the time-evolution of these respective 
density matrices define the Kraus representations, with the following 
expressions for the density matrices for the sub-systems: 
{\scriptsize\begin{eqnarray}
\label{2}
\rho_A(t)={\rm Tr}_B[\rho(t)]&=&\displaystyle\sum_i\, V_i(t)\rho_A(0)V_i^\dag(t),\nonumber \\
 \displaystyle\sum_i\, V^\dag_i(t)V_i(t)&=&I_A,  \\
 \label{3} 
\rho_B(t)={\rm Tr}_A[\rho(t)]&=&\displaystyle\sum_j\, W_j(t)\rho_A(0)W_j^\dag(t)\nonumber \\
 \displaystyle\sum_j\, W^\dag_j(t)W_j(t)&=&I_B.
\end{eqnarray}}
Here  $I_A,\ I_B$ are unit operators in the respective Hilbert spaces of the systems $A$ and $B$. 
The operators $\{V_i(t)\}$, $\{W_j(t)\}$ are expressed in terms of the  matrix elements 
( in the corresponding subsystem spaces) of the unitary time evolution operator $U(t)$  
and will not be made explicit at this juncture. 
The second set of conditions in Eqs. (\ref{2}),(\ref{3}), represent the unit traces of the density matrices. 

Consider a typical Schrodinger operator,  $a_S$ , in the A-space, which in the composite space is represented 
by $a_S\otimes I_B$. Then, 
${\rm Tr}[a_S~\otimes~I_B\, U(t)\rho(0)U^\dag(t)]  
={\rm Tr}_{A}{\rm Tr}_{B}[a_S\otimes~I_B\, U(t)\rho(0)U^\dag(t)]$
 $={\rm Tr}_B[a_S\rho_A(t)]={\rm Tr}_B[\tilde{a}(t)\rho_A(0)],$
and thus the effective time-dependent operator in the space of $A$ 
is 
\begin{equation}
\label{4}
\tilde{a}(t)=\displaystyle\sum_i\, V^\dag_i(t)a_SV_i(t).
\end{equation}
Similarly a typical time-dependent operator in the $B$-space has the form
\begin{equation}
\label{5}
\tilde{b}(t)=\displaystyle\sum_j\, W^\dag_j(t)b_SW_j(t).
\end{equation}
Because these transformations are not unitary, the pristine algebra of the Schrodinger operators is not preserved and %%@
thus
 the dynamical effects of interactions are reflected in these modifications. For example, 
 \begin{eqnarray*} (\tilde{a}(t))^k&\neq& \widetilde{a^k}(t),\\   
 \tilde{a}_{s_1}(t)\tilde{a}_{s_2}(t)&\neq& \widetilde{a_{s_1}a_{s_2}}(t), \ \ \ \ {\rm }etc.
 \end{eqnarray*} 
We now apply this formalism to JCM~\cite{2,3,5} 
and illustrate these features in some detail. 
 
\noindent{\em The JCM Hamiltonian and its exact solution~\cite{2, 5, 11}:} 
We use the convenient notations given in Ref.~\cite{11} with units, where the Planck constant $\hbar=1$ . We present %%@
here an alternate description of JCM in terms of the quantized electromagnetic field (EMF) 
represented by three mutually orthogonal vectors. The direction of propagation of 
photon (say z-direction), electric field perpendicular to the propagation direction, 
say y-direction, and the magnetic field, transverse to both of these along the x-direction. 
In terms of the creation and annihilation operators of the photon field~\cite{12}:
\begin{equation}
\label{6}
E_y=i\sqrt{(\omega/2)}\, (a-a^\dag),\  H_x=\sqrt{(\omega/2)}\, (a+a^\dag) 
\end{equation}
where $\omega$ is the photon frequency (laser field). 
The Hamiltonian of the JCM is then given by:
\begin{eqnarray}
\label{7}
H_{JCM}&=&\frac{E_y^2+H_x^2}{2}\otimes I_A+\frac{\omega_0}{2}\, I_R\otimes\sigma_z\nonumber \\
&&\ \ +\frac{2}{\omega}\, g\, 
(H_x\otimes\sigma_x+E_y\otimes\sigma_y), \nonumber \\
&=& \omega\,(a^\dag a+\frac{1}{2})\otimes I_A+\frac{\omega_0}{2}\, I_R\otimes \sigma_z \nonumber \\
&& \ \ \ \ + g\, (a\otimes \sigma_+ +a^\dag\otimes \sigma_-).
\end{eqnarray}
The first term in Eq.~(\ref{7}) is the Hamiltonian of the radiation with frequency $\omega$. The second term is that %%@
of the atom with 
the energy difference between its ground and excited state $\omega_0$
 and the last term is the dipole interaction between the atom and radiation expressed in the rotating wave %%@
approximation, which leads 
to dynamical, quantum entangled behavior. The second form of the JCM Hamiltonian is the familiar one and opens %%@
interpretation in more general terms of boson fields in place of EMF. Here we employ the standard Pauli matrices and %%@
the photon destruction and creation operators in the number representation with the standard equal time commutation %%@
rule  $[a,a^\dag]=I_R$~\cite{3,11}. 
This Hamiltonian has a constant of the motion, $C=a^\dag a\otimes I_A+\frac{1}{2}\, I_R\otimes \sigma_z$
$=N\otimes I_A+\frac{1}{2}\, I_R\otimes \sigma_z,$ that enables us to find the normal modes 
of the   JCM. The respective unit operators in the radiation and atomic systems are
 $ I_R=\displaystyle\sum_{n=0}^\infty\, |n\rangle\langle n|,\ \ I_A=\vert\uparrow\rangle\langle\uparrow\vert
  +\vert\downarrow\rangle\langle\downarrow\vert.$
  The time evolution operator in terms of the exact solutions is then given by~\cite{11},
  \begin{eqnarray*}
  &&U(t)=\displaystyle\sum_{n=0}^\infty\, \{\,e^{-i\omega\, t(n-\frac{1}{2})}\, v^*_{n-1}(t)\, 
  \vert n, \downarrow\rangle\langle n, \downarrow\vert \\ 
 &&\ \ \ \ \ \ \ \ \ \ \ \ \ \ \ +e^{-i\omega\, t(n+\frac{1}{2})}
  \,[\, v_{n}(t)\,\vert n, \uparrow\rangle\langle n, \uparrow\vert \\
&&  -iw_{n}(t)\,\vert n+1, \downarrow\rangle\langle n, \uparrow\vert 
  -iw_{n}(t)\,\vert n, \uparrow\rangle\langle n+1, \downarrow\vert\,
  ]\,\} 
  \end{eqnarray*}
  with the photon-atom correlation factors
  $v_n(t)=e^{-i\, \lambda_n\,t}\sin^2\theta_n+e^{i\,  \lambda_n\, t}\cos^2\theta_n,$ 
  $w_n(t)=\sin 2\theta_n\, \sin\lambda_n\, t,$ and $\tan\theta_n=\frac{g\sqrt{n+1}}{\left(\frac{\Delta\omega}{2}
  +\lambda_n\right)},$ where $\lambda_n=\left[\left(\frac{\Delta\omega}{2}\right)^2+g^2\, (n+1)\right]^{1/2},$ 
  with detuning frequency $\Delta\omega=\omega-\omega_0.$  
  Consider the initial density matrix to be $\rho(0)=\rho^i_R\otimes\rho^i_A$, where $\rho^i_A$ denotes  $2\times 2$ 
  density matrix of atoms and $\rho^i_R=\vert\alpha\rangle\langle\alpha\vert$  
    represents the pure state of the laser, with  $\vert\alpha\rangle=e^{-|\alpha|^2/2}\displaystyle
	\sum_{n=0}^\infty\frac{\alpha^n}{\sqrt{n}}\,\vert n\rangle,$ $\alpha=|\alpha|\, e^{i\,\phi}$ 
	representing the amplitude and phase of the laser beam. The mean number of photons in the beam is given by 
$M=|\alpha|^2$	  and the phase average of the density matrix gives the mixed Poissonian 
distribution of the photons, $p(n)=\frac{e^{-M}M^n}{n!}.$ The time-dependent atomic marginal 
density matrix is 
\begin{eqnarray}
\label{8}
\rho_A(t)&=&\displaystyle\sum_{N=0}^\infty\, W_{N\alpha}\rho^i_AW_{N\alpha}^\dag\nonumber \\
W_{N\alpha}&=&\langle N\vert U\vert \alpha\rangle,\ W^\dag_{N\alpha}=
\langle N\vert U^\dag\vert \alpha\rangle.
\end{eqnarray} 
  Similarly the time-dependent radiation marginal density matrix is found to be 
  \begin{eqnarray}
  \label{9}
  \rho_R(t)&=&\displaystyle\sum_{s,s_1,s_2=\uparrow,\downarrow}\, (\rho^i_A)_{s_2s_1}\,V_{ss_2}\, 
  \rho^i_{R}\,V^\dag_{s_1s},\ \nonumber \\ 
  V_{ss_2}&=&\langle s\vert U\vert s_2\rangle. 
  \end{eqnarray}
It is important to note that the photon sub-dynamics depends on the initial atomic density matrix and the 
atomic sub-dynamics depends on the initial photon density matrix.
We now express some typical sub-dynamics operators in the Heisenberg representation. 

{\em Photons}: The annihilation operator in the photon sub-dynamics is  
  \begin{eqnarray}
  \label{11}
  \tilde{a}(t)&=&\displaystyle\sum_{s,s_1,s_2} \, (\rho^i_A)_{s_2s_1}\, V^\dag_{s_1s}\, a\, V_{ss_2}\nonumber \\ 
  &=&e^{-i\,t\,\omega}\displaystyle\sum_{n=0}^\infty\left[ |n\rangle\langle n+1|\sqrt{n+1}\, A_n(t)\right.\\
& &\left.  \ + |n-1\rangle\langle n+1|\, C_n(t) +|n\rangle\langle n|\, D_n(t)
  \right],\nonumber
  \end{eqnarray}
  where $A_n(t)=\rho^i_{\uparrow\uparrow}\left(v_n^*(t)v_{n+1}(t)+w_n(t)w_{n+1}(t)\,\sqrt{\frac{n+2}{n+1}}\right)
  +\rho^i_{\downarrow\downarrow}\left(v_n^*(t)v_{n-1}(t)+w_n(t)w_{n-1}(t)\,\sqrt{\frac{n}{n+1}}\right), $\break
  $C_n(t)=i\,\rho^i_{\downarrow\uparrow}\left(w_{n-1}(t)v^*_{n}(t)\, 
\sqrt{n+1}-w_n(t)v^*_{n-1}(t)\sqrt{n}\right)$,\break 
$D_n(t)=i\,\rho^i_{\uparrow\downarrow}\left(  w_{n-1}(t)v_{n}(t)\,\sqrt{n}-w_n(t)v_{n-1}(t)\sqrt{n+1}\right).$
This is a quasi-annihilation operator of the effective photon.  The values $A_n=1,\ C_n=D_n=0$ correspond to the usual %%@
pristine photon 
annihilation operator. The factor $A_n$  represents the suppression of single photon annihilation, while 
$C_n$, $D_n$ represent 2-photon and no-photon annihilations respectively. The precise nature of the photon-QPL will %%@
depend on the choice of the parameters 
and on the initial spin state. The 
standard commutation rule, $[a,a^\dag]=I_R$, does not hold for the photon-subsystem 
operators. 

The number operator, used to construct the number representation (Fock space) in the non-interacting system, takes 
the form
\begin{eqnarray}
\label{12}
\tilde{N}(t)&=&\displaystyle\sum_{s,s_1,s_2} (\rho^i_A)_{s_2s_1}\, V^\dag_{s_1s}\, (a^\dag a)\,V_{s,s_2}
\neq \tilde{a}^\dag(t)\tilde{a}(t)\nonumber \\ 
\tilde{N}(t)&=&\displaystyle\sum_{n=0}^\infty\,\left[\,\vert n\rangle\langle n\vert\, n 
+\vert n\rangle\langle n\vert\, (\rho^i_{\uparrow\uparrow}\, w^2_n(t)-
\rho^i_{\downarrow\downarrow}\, w^2_{n-1}(t))\right.\nonumber \\
&-& i\,\rho^i_{\uparrow\downarrow}\,\vert n+1\rangle\langle n\vert\, w_n(t)v_n(t)\\
&+& \left. i\,\rho^i_{\downarrow\uparrow}\,\vert n\rangle\langle n+1\vert\, w_n(t)v^*_n(t)
\right]. \nonumber 
\end{eqnarray} 
These relations exhibit how the JCM interaction modifies the pristine field operators 
in important ways, thus leading to quasi-number representation in the radiation subsytem.

\noindent{\em Spin:} Using the standard representation of the spin operators we obtain 
the quasi-spin components associated with the atom-subspace,   
$\tilde{\sigma}_k(t)~=~\displaystyle\sum_{N=0}^\infty\,
 W^\dag_{N\alpha}\sigma_k(0)W_{N\alpha}$: 
  \begin{eqnarray}
\label{13}
 \tilde\sigma_+(t)&=& e^{i\omega\, t}\, \left(\vert\uparrow\rangle\langle\downarrow\vert\, S_1^+(t)
 + \vert\downarrow\rangle\langle\uparrow\vert \, S_2^+(t)\right. \nonumber \\
&& \ \ \ \ \ \ \left. + \vert\uparrow\rangle\langle\uparrow\vert\, S_3^+(t) 
 + \vert\downarrow\rangle\langle\uparrow\vert\, S_4^+(t)\right) \\
& =& [\tilde\sigma_-(t)]^\dag, \nonumber
 \end{eqnarray}
 with 
{\scriptsize \begin{eqnarray*}
S_1^+(t)&=& \displaystyle\sum_{n=0}^\infty\,  v^*_n(t)v^*_{n-1}(t)\nonumber \\ 
 S_2^+(t) &=& \displaystyle\sum_{n=0}^\infty\, w_n(t)w_{n-1}(t)\, 
 \frac{\alpha^*\, \sqrt{n}}{\alpha\,   \sqrt{n+1}}\nonumber \\
S_3^+(t)&=& -i\, \displaystyle\sum_{n=0}^\infty\,  v^*_n(t)w_{n-1}(t)\, \frac{\sqrt{n}}{\alpha} \\ 
S_4^+(t)&=&+i\, \displaystyle\sum_{n=0}^\infty\,v^*_{n-1}(t)w_n(t)\, 
\frac{\alpha^*}{\sqrt{n+1}}.
\end{eqnarray*}}
And,
\begin{eqnarray}
\label{14}
 \tilde\sigma_z(t)&=&  \vert\uparrow\rangle\langle\uparrow\vert\, S_1^z(t)
+\vert\downarrow\rangle\langle\downarrow\vert\, S_2^z(t) \nonumber \\
& +&
  \vert\uparrow\rangle\langle\downarrow\vert\, S_3^z(t)+
  \vert\downarrow\rangle\langle\uparrow\vert\, S_4^z(t),  
 \end{eqnarray}
 with
{\scriptsize \begin{eqnarray*}
S_1^z(t)&=& \left[1-2\, \displaystyle\sum_{n=0}^\infty\,p(n)\, 
w^2_n(t)\right]\\
S_2^z(t)&=&-\left[\,1-2\, \displaystyle\sum_{n=0}^\infty\,p(n+1)\, 
w^2_n(t)\,\right]\\ 
S_3^z(t)&=&-2i\,\displaystyle\sum_{n=0}^\infty\,p(n)\, w_n(t)v_n^*(t)\, \frac{\alpha}{\sqrt{n+1}}=S_4^{z*}(t)
 \end{eqnarray*}}
Here again the time-dependent coefficients display the contributions from the JCM interactions 
and they depend on the initial photon state. 
The algebra of these operators is different from that of the pristine counterparts as can be seen directly.  
It is emphasized that these transformations are not unitary, the pristine algebra of 
the Schrodinger operators is not preserved and thus the dynamical effects of interactions are reflected 
in these modifications.

\noindent{\em Results: Explanation of QPL features.} 

The sub-dynamics of the photon and spin are complicated due to the presence of new photon and spin processes as seen %%@
in 
Eqs.~(\ref{11}), (\ref{13}) and (\ref{14}) resulting from JCM. The photon QPL properties will emerge if its %%@
sub-dynamics 
is dominated by $A_n$ in Eq.~(\ref{11}) and the other two terms negligible. Similarly the spin QPL properties 
correspond to a regime, where the first two terms in Eq.~(\ref{14}) dominate, while the last  two terms
 are negligible. Under these circumstances, the sub-dynamics appears to admit 
 the interpretation in terms of fractional numbers of photons and fractional spins. 
  Such QPL regimes emerge when detuning dominates the Rabi term 
 of the coherence factor $\lambda_n,$ i.e.,$\ \Delta\omega>>2g\,\sqrt{n+1}.$ 
 Then the resulting dynamics of QPL photons and spins fall into 
 collapse-revival structure of a large detuned JCM.
In Fig.~1, we display how far the operator structures of the photon, Eqs.~(\ref{11}), (\ref{12}), and the spin, %%@
Eq.~(\ref{14}) 
deviate from their pristine non-interacting versions in the subystem for various values of the system 
parameters.  In the non-interacting case, $\langle\alpha\vert \tilde{a}(t)\vert\alpha\rangle=\alpha$,  
$\langle\alpha\vert \tilde{N}(t)\vert\alpha\rangle=\vert\alpha\vert^2$ 
and two sets of curves in Fig.~(1a) show the effects of interaction on these values as a function of $gt$. 
In Fig.~(1b) we display the eigenvalues of  $\tilde{\sigma}_z(t)$ vs. $gt.$ 
These eigenvalues are given by {\scriptsize$(OS)_z\pm (DS)_z$}, with the off-set  from zero is given by 
{\scriptsize$(OS)_z=(S_1^z(t)+S_2^z(t))/2$} , and the dispersion is given by 
{\scriptsize$(DS)_z=\left\{\left(\frac{(S_1^z(t)-S_2^z(t))}{2}\right)^2+\vert S_3^z\vert^2\right\}^{1/2}.$} 
In the non-interacting case, these are +1 and -1, with off-set equal to 0. The curves in Fig.~(1b) show these 
and, in addition, display the collapse and revivals. The collapses mimic constant values different from $\pm 1,$ 
 depending on the parameters and the revivals occur at different times depending on the detuning. Also, at half the %%@
collapse time, the 
photon figure shows a minimum, and at peaks of revivals, the photons exhibit a maximum. 
We must point out that similar results are obtained when other choices of parameters were made. 
The off-set being small and the dispersion different from +1,-1 and fractional, 
are the results of the non-unitary evolution governing the sub-dynamics. 
A more detailed investigation of these features will be 
explored in another communication.

The back action phenomena may also be interpreted in another 
complementary way. 
We have already shown how the pristine representations of spin and radiation 
get modified due to interaction and entanglement. We now 
exhibit this in terms of the constant of motion, $C$. 
Upon taking the expectation value of this operator over the total 
density matrix of the system, we obtain, 
$\langle n\rangle+\frac{1}{2}\, (\rho^i_{\uparrow\uparrow}-
\rho^i_{\downarrow\downarrow})=\langle\tilde{N}(t)\rangle+\frac{1}{2}\,\langle\tilde\sigma_z(t)\rangle$ , where
$\tilde{N}(t)$ and  $\tilde\sigma_z(t)$ are given explicitly by Eqs. (\ref{12}) and (\ref{14}) 
respectively. The back action on the photon number 
$\langle\tilde{N}(t)\rangle$  is such that it compensates the collapse
and revivals that occur in $\langle\tilde\sigma_z(t)\rangle$. 
Thus the photons and the atom  act 
in tandem in such a way as to maintain the constant of the motion.
\begin{figure}[h]
 \includegraphics*[width=2.4in,keepaspectratio]{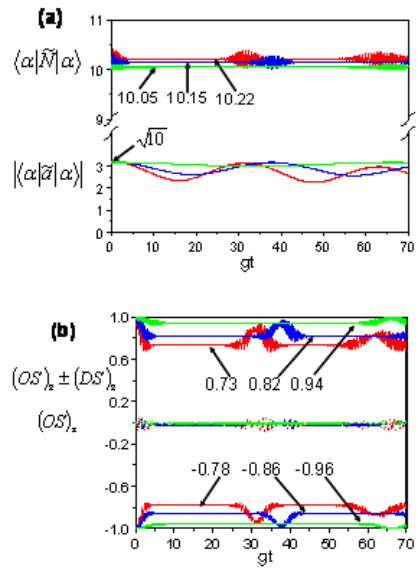}
\scriptsize\caption{\scriptsize(a) Magnitude of $\langle\alpha\vert \tilde{a}(t)\vert\alpha\rangle$  and 
$\langle\alpha\vert \tilde{N}(t)\vert\alpha\rangle$
 vs. $gt$. The set of JCM parameters used are, Atom: $\rho_{\uparrow\uparrow}(0)=1,$
all others zero; Photon:   $N_{\rm mean}=\vert\alpha\vert^2=10;$
Interaction: Red: $\Delta\omega/g=7.5$, 
Blue: $\Delta\omega/g=10$ and Green: $\Delta\omega/g=20$.
Fig.~1\,(b) displays the offsets and eigenvalues of $\tilde{\sigma}_z(t)$ 
vs $gt$ for $N_{\rm mean}=10.$}  
\end{figure}

In conclusion, we have displayed here the complementary dressed behaviors of 
the sub-dynamics of photons (equivalently, the EMF) and atom  in the JCM model of interacting fields. 
We believe that this type of behavior holds generally in the non-unitary sub-dynamic 
evolution  of any bipartite system. The details will differ 
depending on the type of interaction between the two systems.

\end{document}